\documentclass[prl,superscriptaddress,twocolumn,amsmath,amssymb,floatfix]{revtex4-1}
\usepackage{latexsym}
\usepackage{dcolumn}
\usepackage[dvips]{graphicx}
\usepackage{amssymb}
\usepackage{graphics}
\usepackage{amsmath}
\usepackage{epsf}
\usepackage{graphicx}
\usepackage{psfrag, color}
\usepackage{verbatim}

\begin{document}
\title{Strongly metallic electron and hole 2D transport in an ambipolar Si-vacuum field effect transistor}
\author{Binhui Hu}
\affiliation{Laboratory for Physical Sciences, University of Maryland at College Park, College Park, MD 20740}
\affiliation{Joint Quantum Institute, University of Maryland, College Park, Maryland 20742, USA}
\author{M. M. Yazdanpanah}
\affiliation{Laboratory for Physical Sciences, University of Maryland at College Park, College Park, MD 20740}
\affiliation{Joint Quantum Institute, University of Maryland, College Park, Maryland 20742, USA}
\author{B. E. Kane}
\affiliation{Laboratory for Physical Sciences, University of Maryland at College Park, College Park, MD 20740}
\affiliation{Joint Quantum Institute, University of Maryland, College Park, Maryland 20742, USA}
\author{E. H. Hwang}
\affiliation{Joint Quantum Institute, University of Maryland, College Park, Maryland 20742, USA}
\affiliation{Condensed Matter Theory Center, Department of Physics, 
	 University of Maryland,
	 College Park, Maryland  20742-4111}
\affiliation{SKKU Advanced Institute of Nanotechnology and Department of Physics, Sungkyunkwan
  University, Suwon 440-746, Korea } 
\author{S. Das Sarma}
\affiliation{Joint Quantum Institute, University of Maryland, College Park, Maryland 20742, USA}
\affiliation{Condensed Matter Theory Center, Department of Physics, 
	 University of Maryland,
	 College Park, Maryland  20742-4111}

\date{\today}

\begin{abstract}
We report experiment and theory on an ambipolar gate-controlled Si(111)-vacuum field effect transistor (FET) where we study electron and hole (low-temperature 2D) transport in the same device simply by changing the external gate voltage to tune the system from being a 2D electron system at positive gate voltage to a 2D hole system at negative gate voltage. The electron (hole) conductivity manifests strong (moderate) metallic temperature dependence with the conductivity decreasing by a factor of 8 (2) between 0.3 K and 4.2 K with the peak electron mobility ($\sim 18$ m$^2$/Vs) being roughly 20 times larger than the peak hole mobility (in the same sample). Our theory explains the data well using RPA screening of background Coulomb disorder, establishing that the observed metallicity is a direct consequence of the strong temperature dependence of the effective screened disorder.
\end{abstract}

\maketitle

It is now well-established that, quite generically, ``high mobility" and ``low-density" semiconductor-based effectively metallic 2D systems can manifest anomalous low temperature ``metallic" (i.e., $d\sigma/dT <0$ with $\sigma$ being the 2D conductivity) transport behavior, where a modest variation in temperature ($T \approx 0.1K - 4K$) could decrease $\sigma$ by a large amount, with variations in $\sigma(T)$ by as large as a factor of $\sim$2 observed in Si MOSFET based 2D electron systems (2DES) \cite{r1} and GaAs-based 2D hole systems (2DHS) \cite{r2} in a temperature regime ($0.1K-4K$) where phonons are inactive due to the Bloch-Gr\"{u}neisen (BG) suppression of phonon occupancy. This strong metallic temperature dependence (the precise quantitative definition of ``high-mobility" and ``low-density" is materials dependent and varies from system to system \cite{r3}) in 2D semiconductor structures is in sharp contrast with 3D metals where, at low temperatures ($\alt 10K$), the conductivity typically saturates to a disorder-dependent (and temperature-independent) constant ($\sigma_0$) as the system enters the BG phonon scattering regime with $\sigma(T) \approx \sigma_0 - O(T^{4-6})$. 
By contrast, the observed anomalous $\sigma(T)$ in high-mobility and low-density 2D semiconductor systems appears to follow a leading-order linear temperature dependence, with $\sigma(T) \approx \sigma_0 - O(T)$ over a wide temperature range ($0.1K - 4K$) although eventually (for $T<50$ mK) $\sigma(T)$ saturates 
(or manifests weak localization behavior\cite{tracy}), perhaps because of electron heating effects invariably present in semiconductors.

In the current work we report three remarkable new results on the anomalous 2D metallic behavior by combining experiment and theory: (i) we present the first experimental results on the 2D metallic behavior in an ambipolar system 
where the metallic temperature dependence in the conductivity is separately observed for both 2DES and 2DHS in the same device simply by changing an external gate voltage 
(we mention that low-mobility ambipolar Si 2D devices have earlier been reported in the literature\cite{r44} without any observation of the temperature-dependent metallic transport, which is the focus of our study);
(ii) our observed `metallicity' 
(i.e., the temperature-induced fractional change in the conductivity) 
is an unprecedented factor of 8 (2) in the 2DES (2DHS) for $T=0.3-4$ K range and carrier density $\sim 3\times 10^{11}$ cm$^{-2}$ -- this is by far the largest temperature-induced fractional change in the metallic conductivity ever reported in any non-superconducting system in such a small temperature window
-- for example, earlier-studied 2D Si `metallic' systems in the literature \cite{r7} show at most a factor of 3 change in the conductivity in the same temperature window;
(iii) we explain our observations qualitatively by calculating the temperature and density dependent RPA-Boltzmann conductivity using a realistic model of screened Coulomb disorder where the main difference between 2DES and 2DHS arises from the effective valley degeneracy being 6 and 1 respectively by virtue of the qualitatively different band structures in the conduction and the valence band of the Si(111) ambipolar FET structure used in our experiment -- this leads to the 
screened effective disorder in the 2DES being much weaker (and much more strongly temperature-dependent) than in the 2DHS, although both see exactly the same bare disorder, explaining the remarkable difference in the mobility 
and the temperature dependence 
in the two cases.

\begin{figure}[t]
	\centering
	\includegraphics[width=.9\columnwidth]{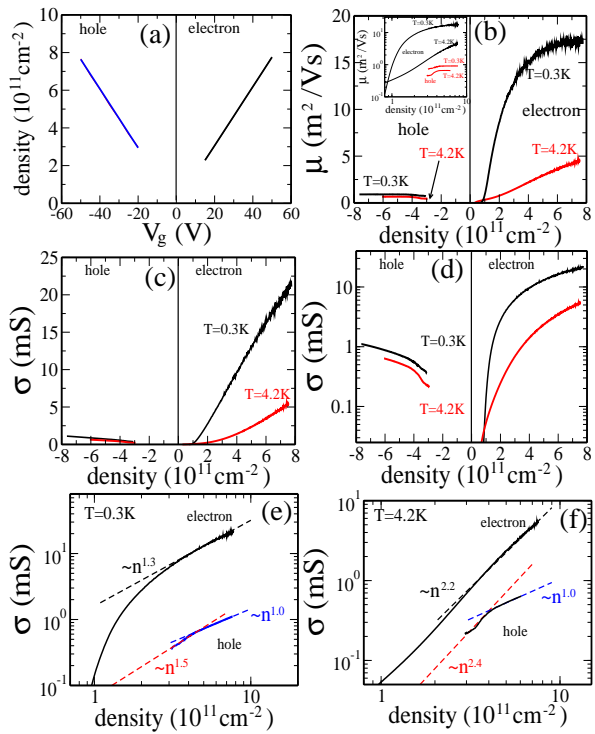}
	\caption{Experimental results for an ambipolar Si(111)-vacuum FET.
(a) Electron and hole densities vs. gate voltage are shown. (b) Mobility 
and (c) conductivity for both electrons (right panel) and holes (left panel) measured at T=0.3K (black) and T=4.2K (red) are shown as a function of density in linear scale. 
Inset shows the same data in (b) as a log plot, where top (bottom) two lines is for electron (hole).
(d) Conductivity vs. density relation is shown in semi-log scale. (e) and (f) show conductivity vs. density in log-log scale at T=0.3K and T=4.2K, respectively. The fitting exponents $\alpha$ in the relation of $\sigma ~\sim n^{\alpha}$ are given in the figures.
}
\label{fig1}
\end{figure}

The ambipolar FET device (we have actually studied several such devices with similar results) we study is a high-purity and high-mobility hydrogen-terminated atomically flat (and nominally undoped) Si(111) structure, where the electrons (holes) are induced near the H-Si(111) surface by applying a positive (negative) voltage through a vacuum barrier. Details of fabrication and  
characterization of such Si-vacuum FETs (in contrast to the usual Si-SiO$_2$ MOSFETs) with ultrahigh mobility, for both electrons \cite{r4} and holes \cite{r5}, have been described elsewhere \cite{r4,r5,r6}. The new aspect of the current work is the fabrication of a Si-vacuum ambipolar FET where we can go from a 2DES to a 2DHS simply by changing the external gate voltage from positive to negative in a single device. 
Such ambipolar devices have earlier been studied for low-mobility Si(100)-SiO$_2$ MOS systems\cite{r44} and for GaAs/AlGaAs based undoped 2D structures \cite{rr1,rr2,rr3,rr4}, but no 2D metallic behavior or temperature-dependent conductivity was reported in either case.
The great advantage of such an ambipolar 
device is that the 2DES and the 2DHS ``feel" precisely the same bare disorder, and therefore a direct comparison between the conductivity data between electrons and holes in the same device should give us considerable insight into the intrinsic aspects of the intriguing metallic 2D phase.

\begin{figure*}[t]
	\centering
	\includegraphics[width=1.8\columnwidth]{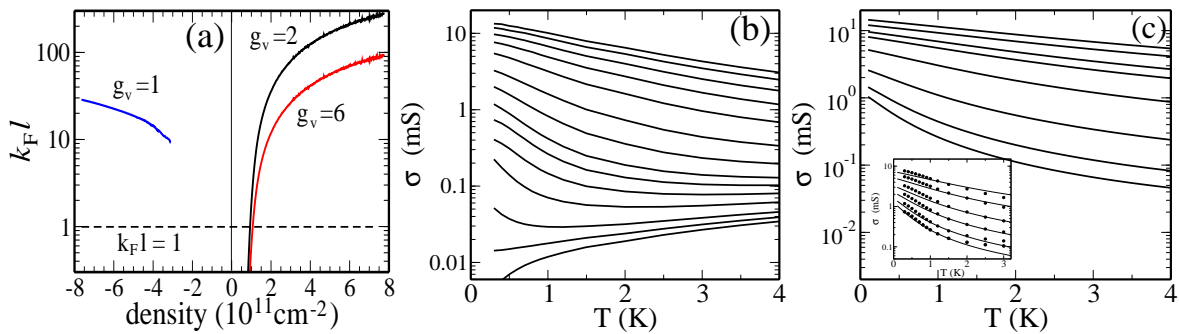}
	\caption{(a) Calculated $k_Fl$ using experimental conductivity with different values of $g_v$. 
(b) The experimentally measured conductivity as a function of temperature for several electron densities, $n=0.85$, 0.91, 0.99, 1.15, 1.30, 1.46, 1.61, 1.92, 2.39, 3.17, 3.94, 4.72, 5.49, 6.12$\times 10^{11}$ cm$^{-2}$ (bottom to top).  (c) Calculated conductivity in the presence of ionized channel impurities and surface roughness 
for electron densities $n=1.3$, 1.5, 2.0, 3.0, 4.0, 4.5, 5.5, 6.0$\times 10^{11}$ cm$^{-2}$ (bottom to top). 
Inset in (c) shows the experiment/theory results together for carrier densities ($n=1.3$, 1.46, 1.61, 1.92, 2.39, 3.17, bottom to top) demonstrating reasonable agreement.}
\label{fig2}
\end{figure*}

In Figs.~\ref{fig1} and \ref{fig2} we show the experimental results for one typical ambipolar device (along with some of our theoretical results to be described below also shown in Fig.~\ref{fig2}(c)). The most important salient features of the experimental results (Figs.~1 and 2) are briefly summarized here: (i) The peak electron (hole) mobility in the device reaches 180,000 cm$^2$/Vs (9,000 cm$^2$/Vs) at $T=0.3$K with an astonishing factor of 20 difference in the electron versus hole mobility
although both are being measured in the same sample at the same temperature and carrier density
-- by contrast, the corresponding high-mobility GaAs/AlGaAs 2D ambipolar devices shows electron and hole mobilities typically within a factor of $2-3$ of each other \cite{rr4} 
which is expected just based on the electron/hole effective mass difference;
(ii) the observed temperature dependence in the conductivity is much stronger for the electrons than for the holes;
(iii) the extrapolation of the mobility (or the conductivity) to low gate voltage indicates a rough mobility gap of 3V, which is approximately the indirect band gap of Si as expected in an 
ambipolar device;
(iv) the carrier 
density dependence of the conductivity, $\sigma(n) \sim n^{\alpha}$ where $n$ is the electron (or hole) density and $\alpha$ is the density exponent of conductivity \cite{r8}, gives $\alpha=1.3$ (1.0) for electrons (holes) at $T=0.3K$ and $\alpha = 2.2$ (1.0) for electrons (holes) at $T=4.2K$ in the `high-density' ($>3\times 10^{11}$ cm$^{-2}$) regime with $\alpha$ increasing at lower density most likely due to density inhomogeneity effects \cite{r9}
which become strong at low carrier density in the presence of random charged impurity centers;
(v) both the conductivity and the mobility (for both electrons and holes) increase monotonically with increasing density with no sign of conductivity saturation (or mobility decrease) at our highest experimental density ($\sim 10^{12}$ cm$^{-2}$), indicating that surface (or interface) roughness scattering, which dominates 2D carrier transport in standard
Si-SiO$_2$ MOSFETs \cite{r10,r11}
and in the GaAs-based gated ambipolar devices \cite{rr3,rr4},
plays (at best) a minor role in the Si-vacuum 2D structures (similar to the corresponding situation in GaAs-based modulation-doped
high-mobility 2D systems \cite{r12}) due to the atomically flat nature of our high-quality Si(111) surface; 
(vi) the main density regime of interest ($>1.5\times 10^{11}$ cm$^{-2}$) for the study of the 2D effective metallic behavior has $k_Fl \gg 1$ (Fig.~2(a)) for both electrons and holes (with $k_F$, $l$ being Fermi wave vector and mean free path, respectively) implying that a Boltzmann theory based transport theory should work well for both the 2DES and the 2DHS existing in our ambipolar device;
(vii) the threshold carrier density (obtained by extrapolating the measured electron or hole Hall density to zero conductivity in Fig.~\ref{fig1}(a)--(d)) is almost the same for the 2DES and the 2DHS with the hole system having only a very small amount of ($\sim 8 \times 10^9$ cm$^{-2}$) higher surface charge states populated by the gate, indicating the very high quality of the sample and that the two systems have almost identical background disorder
(thus indicating that the very large difference in the electron versus hole mobility is an intrinsic effect not arising from any extrinsic difference in the disorder in two cases).

In Fig.~\ref{fig2}(c) we show our theoretically calculated $\sigma(n,T)$ for the 2DES to be compared with
the corresponding experimental data in Fig.~2(b) whereas Fig.~2(a) shows that for density $>10^{11}$ cm$^{-2}$ the Boltzmann theory should be valid as $k_F l \gg 1$ applies for the experimental conductivity. The finite temperature 2D Boltzmann theory has already been described by us in details in our earlier work on Si MOSFETs \cite{r3,r13}, and we only mention that the results shown in Fig.~2(c) use finite-temperature and finite-wave vector RPA screening \cite{r14} of the background disorder which is taken to be unintentional random quenched charged impurity centers in the 2D Si layer itself as well as a small amount of surface roughness. 
The
most important parameter determining the theoretical $\sigma(n,T)$ here is the valley degeneracy ($g_v$) which is taken to be $g_v=6$ consistent with the bulk conduction band structure of six equivalent conduction band minima along the three symmetry axes of Si. Such a high ($g_v=6$) valley degeneracy for the 2DES on the Si(111) surface is consistent with earlier experimental results on high mobility Si-vacuum FETs \cite{r15}, but not with most low mobility Si-SiO$_2$ MOSFET samples studied in the literature \cite{r10} where $g_v=2$ is typically found most likely because of uniaxial interface strain at the Si-SiO$_2$ interface which lifts four of the valleys higher in energy leaving a ground state valley degeneracy of $g_v=2$. 
Our independent SdH analysis of magnetoresistance oscillations (not shown, but see, e.g., Ref.~[\onlinecite{r15}]) in the sample confirms that the system indeed has $g_v=6$.
We emphasize that the effective mass difference between electrons and holes in our Si(111) 2D system (only a factor of 1.67) cannot explain at all the large difference in our measured mobility.

We note that although theory and experiment agree reasonably well qualitatively 
using $g_v=6$
(and even quantitatively for density above $1.3 \times 10^{11}$ cm$^{-2}$)
in Fig.~2, we have not attempted any quantitative fitting because the precise disorder parameters are unknown 
in the experiment.
(We mention that using $g_v=2$ in the theory gives results in qualitative and quantitative disagreement with the experimental data for the 2DES.)

In Fig.~3 we show the theoretical results for two temperatures ($T=0.3$ K and 4.2 K) for both 2DES (both $g_v=2,6$ are shown for the sake of comparison in Fig.~3(a)) and 2DHS (only $g_v=1$ is shown since the Si valence band has no valley degeneracy). The theory reproduces all the key features of the experimental data provided $g_v=6$ (1) is used for the 2DES (2DHS). In particular, there is a very large ($\sim$ a factor of 20) difference in the 2DES and 2DHS mobilities although both see identical disorder. We have checked explicitly that this mobility difference arises mainly from the different valley degeneracies in the two cases -- for example, changing the electron or hole effective mass
does not modify the results much whereas
changing the valley degeneracy for either electrons or holes has a huge effect. 
The theory also reproduces the much stronger temperature dependence of the 2DES conductivity compared with the 2DHS case, again arising primarily from the valley degeneracy difference. Finally, we show in Fig.~3(b) the calculated exponent (with $\sigma \sim n^{\alpha}$) for 2DES and 2DHS at $T=0.3$ K and 4.2 K, finding for the 2DES (with $g_v=6$) the exponent $\alpha = 1.3$ and 2.2 for $T=0.3$ K and 4.2 K,  respectively, and
for the 2DHS (with $g_v=1$) $\alpha=1.0$ and 1.1 for $T=0.3$ K and 4.2 K respectively. These theoretically calculated exponents are in agreement with the experimental data shown in Fig.~1(e) and 1(f). We emphasize that all our theoretical results assume the same bare disorder for both 2DES and 2DHS and incorporate all the realistic 
microscopic details.\cite{r10} 
We note that the small threshold difference of $8 \times 10^9$ cm$^{-2}$ surface charge density 
between the 2DES and the 2DHS
has no quantitative effect on our theoretical results.

\begin{figure}[t]
	\centering
	\includegraphics[width=1.\columnwidth]{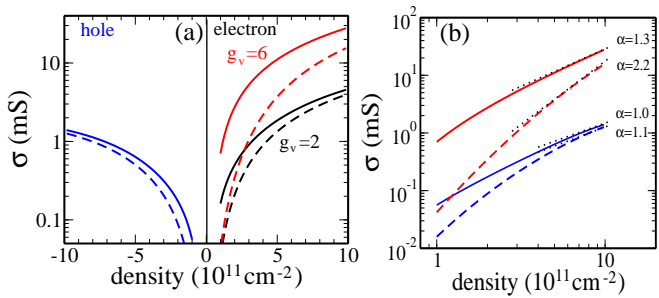}
	\caption{
a) Calculated conductivity as a function of carrier density for two different temperatures $T=0.3$K (solid lines) and 4.2 K (dashed lines).
For the hole system the effective mass $m_h=0.5m_0$ and $g_v=1$ are used, and for the electron system $m_e=0.3m_0$ and $g_v=2$, 6 are used. 
(b) The high density exponents in the relation of $\sigma \sim n^{\alpha}$ are shown for electron system with $g_v=6$ (top two lines) and hole system with $g_v=1$ (bottom two lines). The solid (dashed) lines indicate the calculated conductivity at $T=0.3$ K (4.2 K). The dotted lines are for guide of exponents $\alpha$ shown in the figure.
}
\end{figure}

Before concluding, we provide a simple intuitive understanding of the theory 
which successfully explains the data. At first, the conductivity data appear intriguing because of the huge difference in the quantitive behavior of the conductivity for 2DES and 2DHS in the same sample. Basically, this difference arises from the substantial difference in the effective screened disorder seen by the two kinds of carriers (electrons or holes) in the same ambipolar device because of the large difference in the conduction/valence band structure giving rise to $g_v=6$ (electrons)/1 (holes). The crucial dimensionless quantities \cite{r3,r8} determining both the mobility and the temperature dependence of the conductivity are $q_{TF}/k_F$, where $q_{TF}$ and $k_F$ are the 2D Thomas-Fermi and Fermi  wave vectors, and $T/T_F$, where $T_F$ ($=E_F/k_B$) is the Fermi temperature ($E_F$ is the Fermi energy). We emphasize that the dimensionless interaction strength parameter $r_s \propto m/\sqrt{n}$ is in fact larger for the hole system than the electron system, and is not relevant in controlling the temperature dependence
with the relevant control parameter being $q_s (=q_{TF}/k_F) \sim g_v^{1.5} r_s$ which is much larger for the 2DES compared with the 2DHS in our system.
We assume that only screened Coulomb disorder (and not phonon scattering) determines the conductivity in the 2D Si system as is expected in the $T=0.3- 4.2$ K range.\cite{r10} The constraint on $T/T_F$ is simply 
that it should not be too small in the experimental temperature window for $\sigma(T,n)$ to have strong $T$-dependence. 
It is easy to see that $T_{F}^{(h)}/T_{F}^{(e)} \sim 3$ using $g_v^{(e)} =6$, $g_v^{(h)}=1$ and the respective electron/hole effective masses.  
(The hole effective mass is known to increase from 0.3 to 0.36 in the experimental carrier density range\cite{kot}, but this does not affect our theory in any quantitative manner.)
Thus, the fractional conductivity change, being linear in $T/T_F$ at low temperatures \cite{r14}, is expected to be much larger for 2DES than for 2DHS 
in a given temperature range
simply by virtue of the electron valley degeneracy being six times larger!  But this is only a part of the explanation.
The central quantity of key importance in the theory \cite{r3,r13,r14}
is the dimensionless screening strength $q_s=q_{TF}/k_F$ which determines both the overall magnitude of the mobility as well as the magnitude of the temperature dependence. 
It is easy to see that $q_s^{(e)}/q_s^{(h)}= (m_e/m_h)(g_v^{(e)}/g_v^{(h)})^{1.5} \sim 8$, which implies that the effective screening is much stronger for the 2DES than for the 2DHS, leading to the conclusion that the mobility ratio for the 2DES compared with 2DHS goes approximately as $(m_e/m_h) (q_s^{(e)}/q_s^{(h)})^2 \sim 35$, whereas the exact numerical calculation gives more a factor of 20 difference since the system is not strictly in the $q_s \gg 1$ and/or $T/T_F \ll 1$ limit that these analytical approximations assume.  Similarly, the simplest analytical theory predicts that the temperature-induced fractional conductivity change should go as $(q_s^{(e)}/q_s^{(h)})(T_F^{(h)}/T_F^{(e)})  \sim  20$ assuming that $q_s \gg 1$ and $T/T_F \ll 1$ for both 2DES and 2DHS.  Since these strong-screening and low-temperature conditions are not obeyed in the experiment, the realistic difference in the temperature-dependent conductivity, as obtained in our numerical results, is around a factor of 4.
For the theoretical details we refer to the existing literature.\cite{r3,r13,r14}

In conclusion, we report the first experimental observation of very strong metallic temperature dependence of 2D conductivity in both electrons and holes in an ambipolar Si(111) system, with the electron (hole) conductivity changing by a factor of 8 (2) at a density of $3\times 10^{11}$ cm$^{-2}$ for a temperature change from 0.3 K to 4.2 K with the electron mobility
being 20 times larger than the hole mobility. 
We provide a theoretical explanation for the data using an RPA-Boltzmann transport theory assuming background screened Coulomb disorder as the primary scattering mechanism. Our work conclusively
shows the dominant role of valley degeneracy in determining 2D transport through carrier screening of Coulomb disorder and explains the main difference between the electron and the hole conductivity as arising from the factor of six difference in their valley degeneracy. In particular, we find that the dimensionless parameter $q_{TF}/k_F$
and not the so-called $r_s$-parameter with $r_s \sim m/\sqrt{n}$ controls the strength of metallicity in the anomalous 2D metallic phase of semiconductor systems.

This work is supported by LPS-CMTC. Part of this work was performed at the NIST Center for Nanoscale Science and Technology, and the support of the Maryland NanoCenter through its FabLab is also acknowledged.

\end{document}